\documentclass[preprint,aps]{revtex4}
\usepackage{graphicx}
\usepackage{booktabs}
\usepackage[usenames, dvipsnames]{color}
\usepackage{multirow}
\usepackage{float}
\usepackage{eurosym}
\usepackage{setspace}
\usepackage{gensymb}
\usepackage{rotating}
\usepackage[ pdftex, plainpages = false, pdfpagelabels, 
                 pdfpagelayout = useoutlines,
                 bookmarks,
                 bookmarksopen = true,
                 bookmarksnumbered = true,
                 breaklinks = true,
                 linktocpage=all,
                 pagebackref=false,
                 colorlinks = true,
                 linkcolor = BrickRed,
                 urlcolor  = blue,
                 citecolor = BrickRed,
                 anchorcolor = green,
                 hyperindex = true,
                 hyperfigures
                 ]{hyperref}

\newcounter{figref}

\begin{document}

\title{\href{http://www.necsi.edu/research/social/newyork}{Sentiment in New York City: A High Resolution Spatial and Temporal View}} 
\date{August 20, 2013}  
\author{Karla Z. Bertrand, Maya Bialik, Kawandeep Virdee, Andreas Gros and \href{http://necsi.edu/faculty/bar-yam.html}{Yaneer Bar-Yam}}
\affiliation{\href{http://www.necsi.edu}{New England Complex Systems Institute} \\ 
238 Main St. S319 Cambridge MA 02142, USA}

\begin{abstract}
Measuring public sentiment is a key task for researchers and
policymakers alike. The explosion of available social media data allows for a more
time-sensitive and geographically specific analysis than ever
before. In this paper we analyze data from the micro-blogging site
Twitter and generate a sentiment map of New York City. We
develop a classifier specifically tuned for 140-character Twitter
messages, or tweets, using key words, phrases and emoticons to
determine the mood of each tweet. This method, combined with
geotagging provided by users, enables us to gauge public sentiment on
 extremely fine-grained spatial and temporal scales. We find that
public mood is generally highest in public parks and lowest at transportation hubs, and locate other areas of strong sentiment such as cemeteries, medical centers, a jail, and a sewage facility. Sentiment progressively improves with proximity to Times
Square. Periodic patterns of sentiment
fluctuate on both a daily and a weekly scale: more positive tweets
are posted on weekends than on weekdays, with a daily peak in
sentiment around midnight and a nadir between 9:00 a.m. and noon.
  \end{abstract}

\maketitle

Twitter is a microblogging site created in 2006 \cite{TwitterBlog} that is used by over
500 million
people worldwide \cite{TechCrunch}. Researchers have found it an invaluable data repository for
opinion mining and prediction in a number of fields, including
politics \cite
{Diakopoulos2010,Tumasjan2010,Bermingham2011} 
 and financial markets \cite{Mittal2011, Bollen2011JCS,
Bollen2011Computer}.
Twitter has also served as a resource for predicting and tracking
the propagation of natural disasters, epidemics, and
terrorist incidents \cite{Doan2011}. 
Public mood can be
quantified using Twitter and other Internet sources \cite{Facebook2010, Facebook2011, Kramer2010,Mishne2006,Mishne2005,Yamashita2010}. The usefulness of
Twitter as a tool for sentiment analysis has been
verified by comparison to more traditional metrics,
such as polls \cite{OConnor2010}, socio-economic conditions
\cite{Quercia2012}, and stock market performance \cite{Sharma2010}, as
well as common sense predictors such as weather \cite{Hannak2012}. Recent studies have begun to explore both the dynamics \cite{Thelwall} and geography \cite{Mitchell} of twitter sentiment. 

Here we use Twitter to study the fine-grained geography and dynamics of sentiment in the
greater New York City area, identifying areas and times of positive and negative sentiment. We collected 603,954 tweets via Twitter's API \cite{TwitterAPI}, restricted to
those which were tagged with geocoordinates 
around the immediate New York metropolitan area over the course of two
weeks in April 2012. We built a sentiment classifier in order to assess
the mood of the tweets. Using emoticons we constructed a customized classifier directly from tweets rather than lexicons obtained from other sources. A set of tweets that included
“emoticons” such as \textbf{;)}  or \textbf{:-(} served as the training corpus for positive and negative sentiment classifiers. Then, for each tweet in our full set, we removed the URLs and usernames,
tokenized the text, and assigned a sentiment score based on these two
classifiers. Methodological details are in the Appendix. 

\begin{figure}[h]
\refstepcounter{figref}\label{fig:map}
\includegraphics[width=1\linewidth]{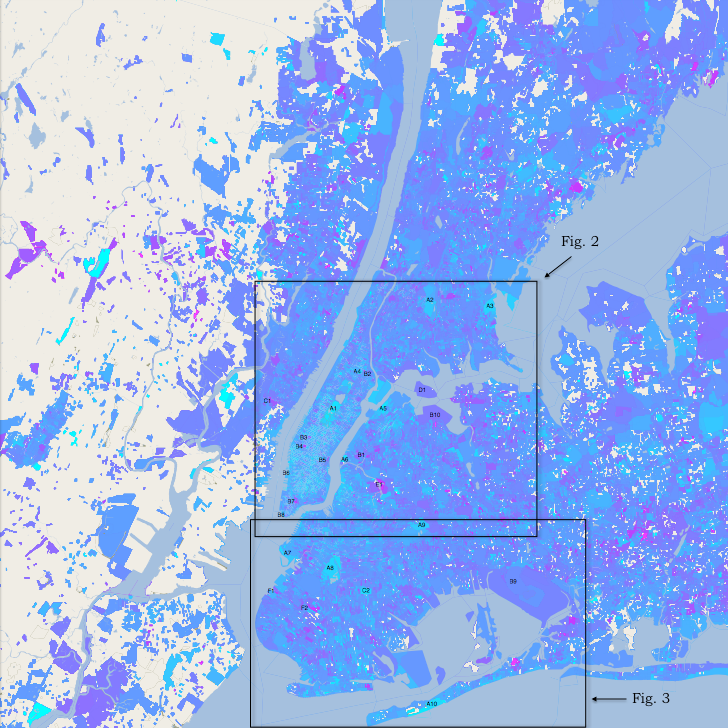}
\caption{Public sentiment map of the New York City
  metropolitan area according to analysis of over 600,000
  tweets, organized by census block. Cyan represents the most positive sentiment and
  magenta the most negative. White represents areas with insufficient tweet density for analysis. The region analyzed consists of the area between latitudes 40\degree~to 41\degree
N and longitudes 73\degree~to 74\degree W. The boxes indicate subareas depicted in subsequent figures.}
\end{figure}

\begin{figure}[h]
\refstepcounter{figref}\label{fig:mapnorth}
\includegraphics[width=1\linewidth]{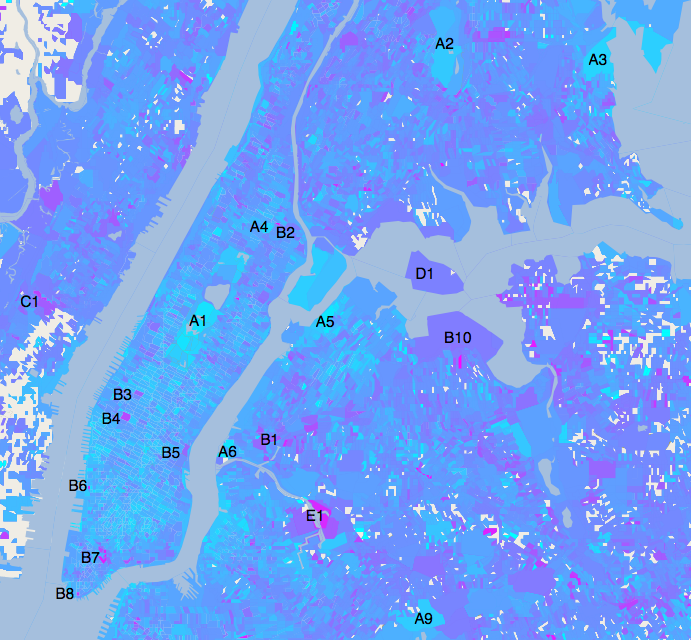}
\caption{Public sentiment analysis of Manhattan and surrounding areas. Areas of strong sentiment are labeled by A -- F as follows: A: Parks; B: Transportation Hubs; C: Cemeteries; D: Riker's Island ; E: Maspeth Creek; F: Medical Centers. For a full list, see Table ~\ref{table:areasonmap}.}
\end{figure}

\begin{figure}[h]
\refstepcounter{figref}\label{fig:mapsouth}
\includegraphics[width=1\linewidth]{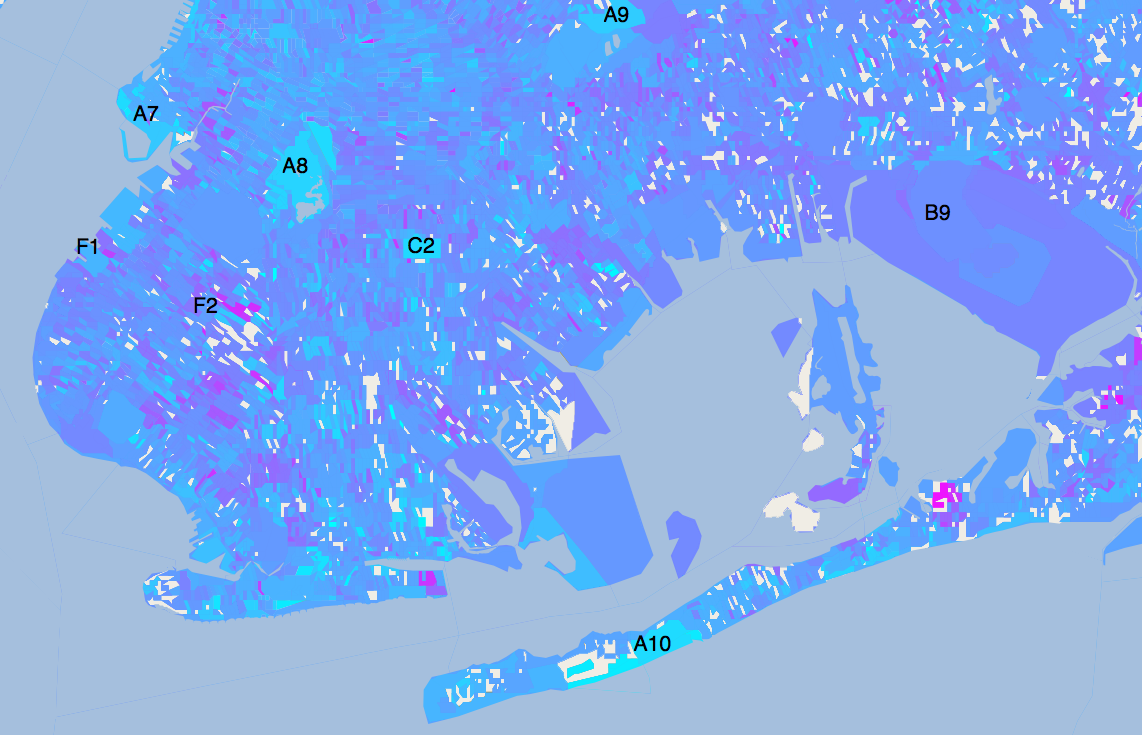}
\caption{Public sentiment analysis of Brooklyn and surrounding areas. Areas of strong sentiment are labeled by A -- F as follows: A: Parks; B: Transportation Hubs; C: Cemeteries; D: Riker's Island ; E: Maspeth Creek; F: Medical Centers. For a full list, see Table ~\ref{table:areasonmap}.}
\end{figure}

We combined the sentiment ratings with geotags to create
a map of public mood, as shown in
Fig.~\ref{fig:map}. Expanded views are shown in Fig.~\ref{fig:mapnorth}, centered on the island of
Manhattan, and Fig.~\ref{fig:mapsouth}, centered on Brooklyn. 

By comparing the sentiment map to a street map of New York, we determine the landmarks corresponding to areas with high and low sentiment (figure labels {\bf A} through {\bf F} in Table ~\ref{table:areasonmap}). Generally, areas of very positive sentiment are public parks and gardens ({\bf A1 -- A10}). This is consistent with knowledge from urban design \cite{Garvin2011}, as well as increasing empirical evidence of the importance of public parks to residents' quality of life \cite{Chiesura2004}. Central Park ({\bf A1}) is visible as a long rectangle of lighter cyan running down the center of Manhattan.

Many of the areas with very negative sentiment are related to transportation hubs such as entrances to tunnels and bridges ({\bf B1 -- B10}), including the Midtown Tunnel ({\bf B5}), the Brooklyn Bridge ({\bf B7}), Port Authority Bus Terminal ({\bf B3}) and Penn Station ({\bf B4}). This is consistent with the finding that people get angry in traffic \cite{DeffenbacherEtAl2002}. Difficulties with other forms of transportation such as buses, trains and flights also seem to be sources of negative sentiment.
\begin{table}[h]
\centering
\caption{Areas of Strong Sentiment}
{\small
\begin{tabular}{|l|}
\hline
{\bf Parks}\\ 
A1. Central Park\\
A2. New York Botanical Garden\\
A3. Pelham Bay Park\\
A4. Marcus Garvey Park\\
A5. Astoria Park\\
A6. Gantry Plaza State Park\\
A7. Red Hook Park\\
A8. Prospect Park   \\
A9. Highland Park   \\
A10. Jacob Riis Park   \\ \hline
{\bf Bus and Train Stations }  \\ 
B1. Hunterspoint Avenue train station  \\
B2. Harlem 125th St. station  \\ 
B3. Port Authority Bus Terminal \\
B4. Penn Station   \\ \hline
{\bf Entrances to Bridges and Tunnels }  \\ 
B5. Midtown Tunnel   \\ 
B6. Holland Tunnel   \\
B7. Brooklyn Bridge   \\
B8. Hugh L. Carey Tunnel   \\ \hline
{\bf Airports }  \\ 
B9. JFK Airport  \\
B10. LaGuardia Airport   \\ \hline
{\bf Cemeteries }  \\
C1. Palisades Cemetery / Weehawken Cemetery   \\
C2. Holy Cross Cemetery   \\ \hline
{\bf Jail Complex }  \\ 
D1. Riker's Island   \\ \hline
{\bf Sewage Facility }  \\ 
E1. Maspeth Creek   \\ \hline
{\bf Medical Centers}  \\ 
F1. Lutheran Medical Center   \\
F2. Maimonides Medical Center   \\ \hline
\end{tabular}
}
\label{table:areasonmap}
\end{table}

\begin{figure}[tb]
\refstepcounter{figref}\label{fig:spatial}
\href{http://www.necsi.edu/research/spatial.pdf}{\includegraphics[width=1\linewidth]{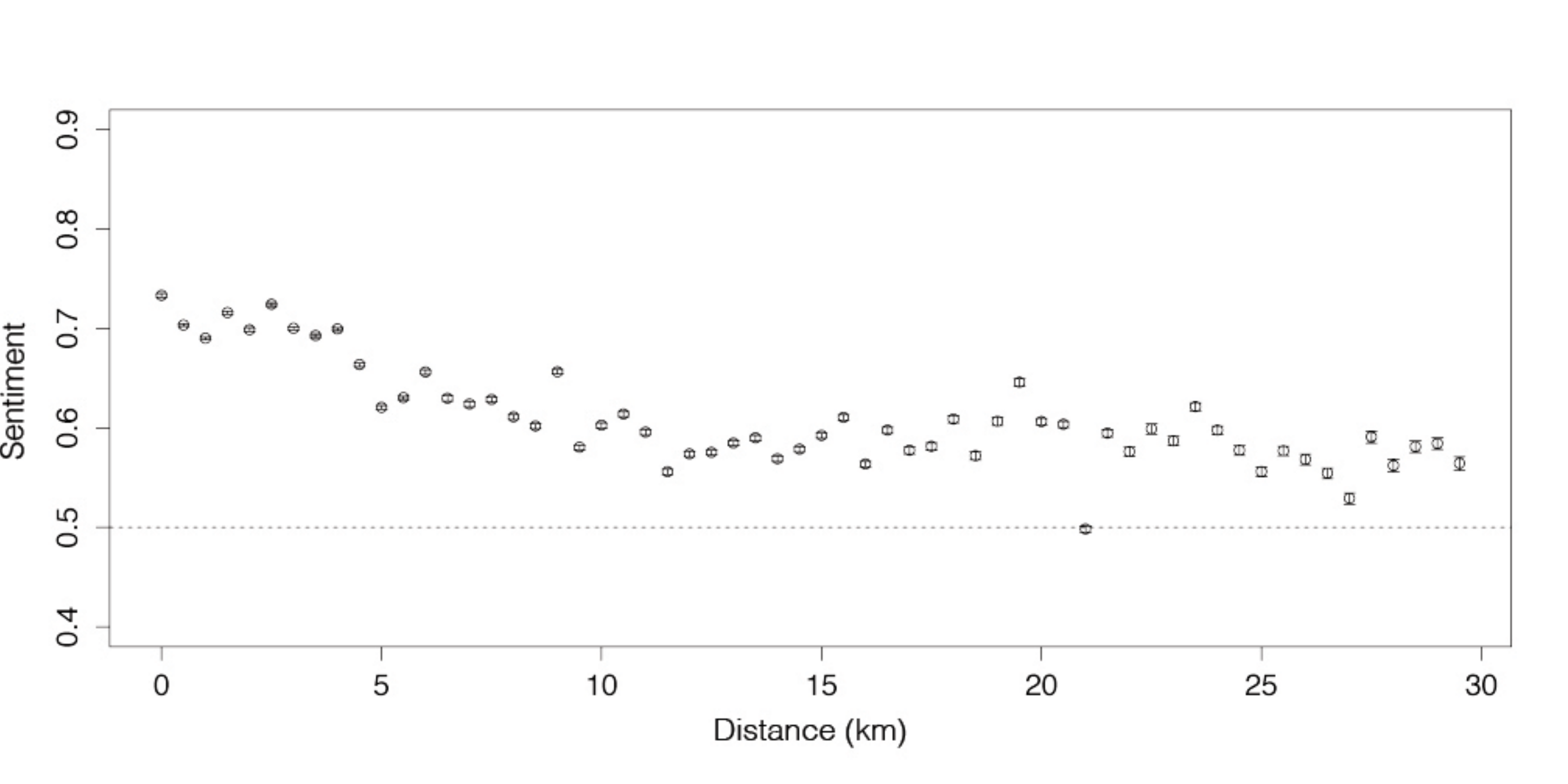}}
\caption{\textbf{Spatial analysis} -- Public sentiment as reflected by
  proportion of positive and negative tweets, decreases as a function of distance from Times
  Square (km).}
\end{figure}

\begin{figure}[tb]
\refstepcounter{figref}\label{fig:temporal}
\href{http://www.necsi.edu/research/temporal.png}{\includegraphics[width=1\linewidth]{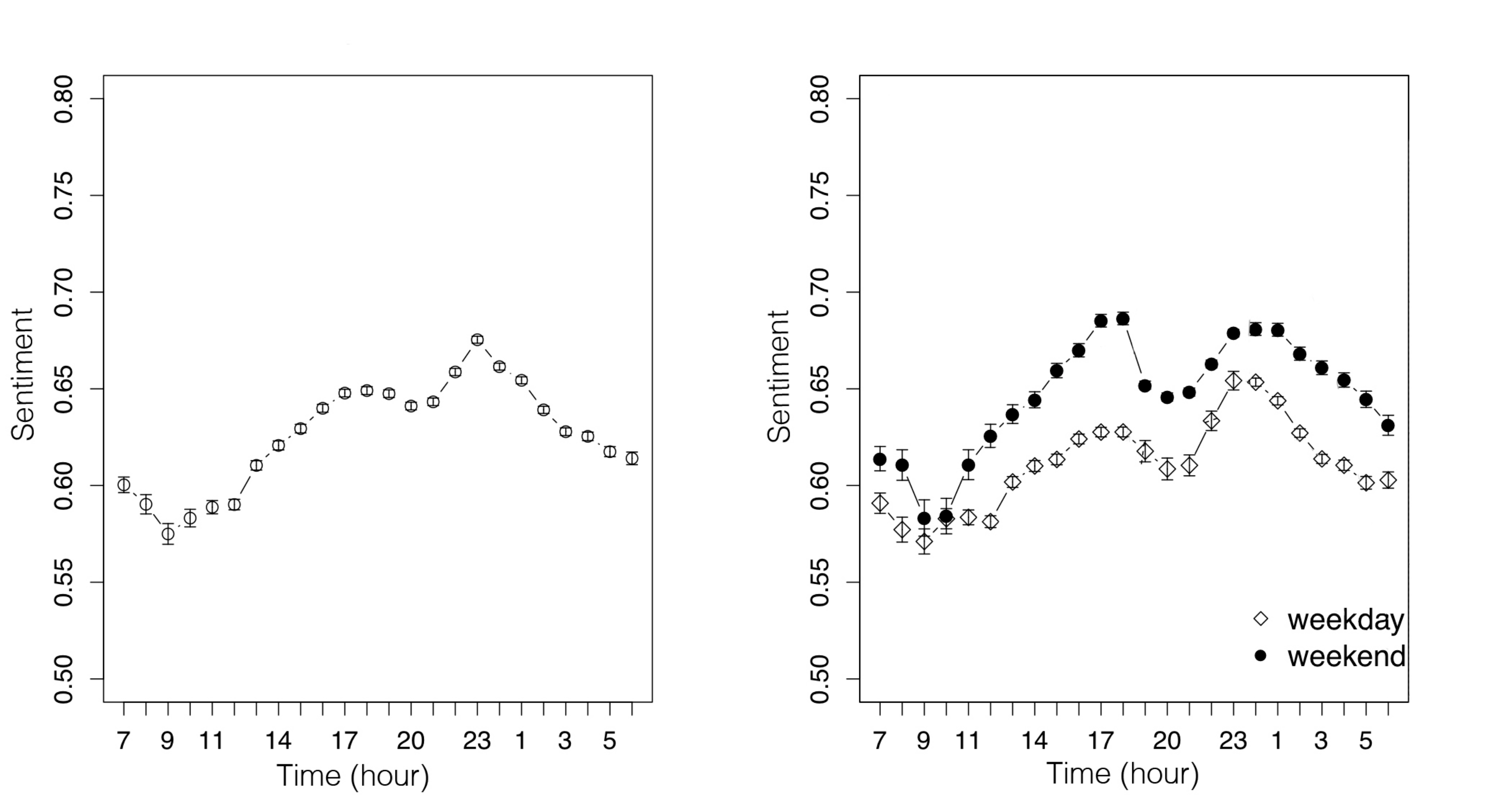}}
\caption{\textbf{Temporal analysis} -- Public sentiment over the course of the
  day. \textbf{Left:} All days. \textbf{Right:}
  Weekdays (open diamonds) and weekend days (filled circles) plotted
  separately.} Error bars represent the standard error of the mean.
\end{figure}
Some cemeteries show strong sentiments: the Palisades and Weehawken cemeteries ({\bf C1}) are negative and the Holy Cross cemetery ({\bf C2}) is positive. It is unclear why the sentiment at the Holy Cross cemetery is positive. Medical Centers ({\bf F1 -- F2}) and Riker's Island ({\bf D1}) --- New York City's main jail complex --- are also areas of strong negative sentiment. 

One area with markedly negative sentiment is Maspeth Creek in Brooklyn ({\bf E1}). While its geographic features are unremarkable, this area is ``one of the most polluted urban water bodies in the country,'' according to EPA regional administrator Judith Enck \cite{HuffPost}. Once one of New York's industrial hubs with more than 50 refineries along its coast \cite{EPA}, this area was the site of the largest oil spill in the country \cite{MaspethVideo} and now contains a 15-foot-thick layer of petroleum-based pollutants that scientists have dubbed ``black mayonnaise'' \cite{HuffPost}. The creek continues to receive 288 million gallons of untreated sewage a year \cite{NewtownPentacle} and was listed as a Superfund site on the National Priorities List in 2010 \cite{EPA}. Although there is a clean up plan in effect and 12 million gallons of oil have already been extracted \cite{HuffPost} and resold \cite{MaspethVideo}, current efforts to meet the standards for oxygen levels have involved aerating the water, and by doing so, spreading the bacteria into the air \cite{dueker2012local}. The Hudson River program director Phillip Musegaas claims ``if and how it affects the local population is somewhat ... obscure'' \cite{MaspethVideo}; our findings of negative sentiment reflect the impact on the local population. 

In Fig. \ref{fig:spatial} we plot the proportion of positive
tweets as a function of distance from Times Square. The sentiment declines from a highly positive value, ~0.74, to ~0.6. The drop is not solely a property of Times Square itself, as the distance until the sentiment declines to a steady level is approximately 10km. The trend
towards more positive tweets nearer to Times Square may be due in part to tourism at famous locations and landmarks in the heart of Manhattan, as well as proximity to Central Park.

The temporal patterns of fluctuations in sentiment shown in Fig. \ref{fig:temporal} reflect the common notion that people are generally happier when they are not at work. On a weekly scale, sentiment is consistently more positive on the weekends than on the weekdays, though the dynamics over the span of one day are remarkably similar in both cases. On a daily scale, sentiment is low in the mornings at 9:00 am, presumably when many people are arriving at work, and steadily rises until it reaches a peak at approximately 4:00-5:00 pm, when many finish their workday. After work there is another rise of sentiment, with a peak at approximately midnight and a monotonic decrease through the night until 9:00 am.

In summary, we have analyzed the geographical and temporal sentiment in New York City during a period of two weeks in April 2011. By building classifiers based on emoticons and averaging sentiment across census blocks, we are able to map the general trends and identify areas of strong sentiment. We find high sentiment in parks and low sentiment in areas of transportation such as train stations, bus stations, entrances to bridges and tunnels, and airports. Other areas of strong sentiment include cemeteries, medical centers, a jail complex, and a polluted sewage facility. We also find a temporal pattern of fluctuation during the weekdays and a similar but higher pattern of fluctuation during the weekends. 

Our method of public mood analysis has several strengths. By
utilizing Twitter's abundance of geotagged data, we can obtain spatial
information that is both wide-ranging and fine-grained. The brevity of tweets allows for rapid processing and
classification, while their frequency produces a time-sensitive
picture of public sentiment. Because we do not use semantic analysis
or lemmatization in our classification, our method is language
independent--- a significant advantage given that fewer than 40\% of tweets
worldwide are composed in English
\cite{http://www.mediabistro.com/alltwitter/twitter-language-share_b16109} and that previous studies of twitter sentiment have been constrained by lexicons which were not adapted to twitter data \cite{Mitchell}.

\section{Appendix}

We collected tweets from the Twitter Streaming Application
Programming Interface (API) \cite{TwitterAPI}, filtering by location to
those geotagged within a bounding box of New York City. We retrieved a
total of 603,954 of these tweets between the latitudes of 40\degree~to 41\degree
N and longitudes of 73\degree~to 74\degree W, from April 13th--April 26th, 2012.

We constructed a classifier based upon supervised learning \cite{NLTK}  with
training labels determined by the presence and type of the emoticons in Table ~\ref{table:emoticons}.

\begin{table}[H]
\centering
\caption{Emoticons used in training of classifiers.}
    \begin{tabular}{|c|c|}
\hline
Positive Emoticons & Negative Emoticons \\ \hline
:) & :( \\ 
:-)� & :-(� \\ 
:D & ;( \\ 
=) & =( \\ 
=D & : (� \\ 
;) & :/ \\ 
\textless3 & :-/� \\ 
:p & ): \\ 
:-p� & ~  \\ 
(: & ~   
    \\ \hline
\end{tabular}
\label{table:emoticons}
\end{table}

For each tweet in this corpus we check if the tweet contains an emoticon.  Each tweet receives two binary labels: true/false on containing a positive emoticon and true/false on
containing a negative emoticon.  We use these labeled tweets to build two
classifiers, one to classify whether or not a tweet is positive, and another to
classify whether or not a tweet is negative.  

To implement the classifiers we use the Natural Language Toolkit (NLTK) 
\cite{NLTK}. Each tweet is first standardized: emoticons are removed,
any urls are replaced by ``URL'', and any username is replaced by ``USER''.  
We tokenize the tweet and treat each unique
token within a tweet as a feature. We use two Bayes classifiers, training one
on the true/false positive-labeled feature sets, and another on the true/false
negative-labeled feature sets. Using both classifiers, we generate a
sentiment measure by combining positive and negative values as in
\begin{math} \frac{p_1 + (1-p_2)}{2} \end{math}. Human checkers were used to verify the reasonableness of a sample of tweets.

\end{document}